\documentclass[hidelinks,12pt]{article}
\usepackage{a4}
\usepackage{moreverb,relsize,url}

\usepackage{pslatex}
\usepackage{hyperref}

\usepackage{graphicx}
\usepackage[listings,skins]{tcolorbox}

\definecolor{mygreen}{rgb}{0,0.1,0}
\definecolor{mygray}{rgb}{0.5,0.5,0.5}
\definecolor{mymauve}{rgb}{0.58,0,0.82}

\usepackage[utf8]{inputenc}
\usepackage{pdflscape}
\usepackage{afterpage}

\usepackage{cite}

\hypersetup{%
   citecolor=black
}

\usepackage{array}
\usepackage{tabularx}

\begin{document}

\title{A Newcomer In The PGAS World - UPC++ vs UPC: A Comparative Study}

\author{
Jérémie Lagravière$^{1}$
\and
Johannes Langguth$^{1}$
\and
Martina Prugger $^{2}$
\and
Phuong H. Ha$^{3}$
\and
Xing Cai$^{1,4}$
}

\date{
{\small $^{1}$Simula Research Laboratory, P.O.~Box 134, NO-1325 Lysaker, Norway}\\
{\small $^{2}$ Lopez Laboratory, Vanderbilt University
 , Nashville, Tennessee, USA.}
{\small $^{3}$The Arctic University of Norway, NO-9037 Troms{\o}, Norway}\\
{\small $^{4}$University of Oslo, NO-0316 Oslo, Norway}\\
\vspace{.5cm}
{\scriptsize Correspondence should be addressed to Jérémie Lagravière; jeremie.lagraviere@gmail.com}
}

\maketitle

\begin{abstract}
A newcomer in the Partitioned Global Address Space (PGAS) 'world' has arrived in its version 1.0: Unified Parallel C++ (UPC++). UPC++ targets distributed data structures where communication is irregular or fine-grained. The key abstractions are global pointers, asynchronous programming via RPC, futures and promises. 
UPC++ API for moving non-contiguous data and handling memories with different optimal access
methods resemble those used in modern C++. In this study we provide two kernels implemented in UPC++: a sparse-matrix vector multiplication (SpMV) as part of a Partial-Differential Equation solver, and an implementation of the Heat Equation on a 2D-domain. Code listings of these two kernels are available in the article in order to show the differences in programming style between UPC and UPC++. We provide a performance comparison between UPC and UPC++ using single-node, multi-node hardware and many-core hardware (Intel Xeon Phi Knight's Landing).
\end{abstract}

{\bf Keywords}: {PGAS; APGAS;UPC++; UPC programming language; Fine-grained irregular communication; Sparse matrix-vector multiplication; Performance optimization}

\section{Introduction \& Motivation}
\label{intro}
In distributed memory parallel systems, MPI has been the \textit{de-facto} standard for a long time. It reliably provides high communication performance, but programming MPI applications is very complex, and this complexity of parallel programming is a key challenge that the HPC research and industry communities face. One of the most prominent approaches to alleviate this problem is the use of Partitioned Global Address Space (PGAS) systems.
For more than 20 years \cite{UPCPointOfOrigin,coArrayFortranPointOfOrigin}, the dominating PGAS implementations have been UPC, Coarray Fortran, and SHMEM. Other implementations have also been proposed such as Chapel \cite{chapelPointOfOrigin}, Titanium, and recently UPC++ \cite{upcxxchangelog}, whose version 1.0 was released in September 2019.  The PGAS programming model constitutes an alternative \cite{PGASAsAnAlternative} to using MPI
or MPI with OpenMP. In this study we focus on UPC++.

We have multiple goals for this study: we want to extend our previous studies \cite{lagraviere2016performance,lagraviere2019performance} and focus on what could be seen as the future of PGAS by studying the new Version 1.0 of UPC++ PGAS language \cite{upcxxchangelog}), released in September 2019. 
To this end, we compare the performance and programability of UPC and UPC++.

PGAS~\cite{1263470,Almasi2011,PGAS_survey,PGAS_site} is a programming model that aims to achieve both good programmer productivity and high computational performance. These goals are usually conflicting in the context of developing parallel code for scientific computations. The main aspect of PGAS is a global address space that is shared among concurrent processes, running on different nodes of a supercomputer 
that jointly execute a parallel program. 
Data exchange between the processes is typically performed transparently by a low-level network layer such as GASNet \cite{gasnet-lcpc18}, 
 without explicit involvement from the programmer, thus providing good productivity.
The shared global address space is logically partitioned 
such that each partition has affinity to a designated owner process. This awareness of 
data locality is essential for achieving good performance of parallel
programs written in the PGAS model, because the globally shared
address space may encompass many physically distributed memory sub-systems.

As mentioned before, UPC and UPC++ are both implementations of the PGAS programming model. UPC stands for {\em Unified Parallel C} and UPC++ stands for {\em Unified Parallel C++}. In both cases data is either {\tt shared} or {\tt private}, where shared data is accessible (read or write) by all threads and private data is only accessible by its owner thread.  UPC is based on the idea of "communication simplification" in  the sense that there is no way for the programmer to distinguish intra-node memory operations (between threads running on the same hardware node) from their inter-node counterparts. This brings a very efficient and simple way of creating programs either in UPC at the very possible expense of performance. By default, UPC++ follows the same pattern, but with the recent addition of {\em teams} of threads, it offers a native way to distinguish between intra- and inter-node communication.  

However, the main difference between UPC and UPC++, on a conceptual level, is that UPC++ implements a sub-category of PGAS, which is called APGAS: "Asynchronous Partitioned Global Address Space". The APGAS model extends the PGAS
with ideas from task-based asynchronous execution model, which describes the semantics of an application as a hierarchy of tasks that are dynamically created and scheduled during the execution time \cite{martin2018mate}. Another language that implements the APGAS paradigm is X10 \cite{X10}. Furthermore, for software maintenance reasons, UPC++ is distributed as a library rather than a language, although this has little effect on the actual PGAS programming.

To study UPC++ we have developed two applications based on well-known kernels: the heat equation in 2D and a sparse-matrix vector multiplication using the ELLPACK format. It is designed to simulate diffusion processes over unstructured 3D mesh representing the human cardiac ventricle. 
In this paper we compare UPC++ to its ancestor UPC on both multi-node and many-core architectures using multiple criteria such as: programmability, performance predictability, scalability, performance (GFLOPS or execution time). After this comparison we discuss the obtained results.

Previous studies \cite{bachan2019upc++,6877339}, made in part by the UPC++ design team, have included comparative aspects about UPC++.  Our study focuses on comparing UPC++ to UPC by using the SpMV kernel which, due to its irregularity and communication requirements, poses a challenging benchmark for distributed memory systems. 


\section{Implementations}
\label{implementations}
\subsection{Kernels}
In this study we have focused our efforts on the implementations of two kernels in UPC and UPC++.
The first kernel is rather simple and is the implementation of the Heat Equation in two dimensions domain, presented in detail in Section \ref{heatEquationKernel}.
The second kernel is the implementation of a Sparse matrix-vector multiplication, solving a 3D diffusion equation that is posed on an irregular domain modeling the left cardiac ventricle of a healthy male human. This 3D diffusion solver was designed as an integral part of a cardiac electrophysiology simulator and is presented in detail in Section \ref{spmvKernel}.
This section aims also at showing the difference between the programming 'style' of UPC and UPC++.

\subsubsection{Heat Equation}
\label{heatEquationKernel}
The heat equation is one of the most fundamental kernels in scientific computing. Due to its simplicity, it is useful as a benchmark to assess the performance of a parallel system. For this experiment, we solve the 2D heat diffusion equation:

$$\frac{\partial \phi}{\partial t}=\frac{\partial^2\phi}{\partial x^2}+\frac{\partial^2\phi}{\partial y^2}$$ 

on a uniform mesh. We employ the finite difference discretization given by:

$$ u(x,y) = \frac{u-h}{2h}$$

We use an existing UPC code\footnote{The code was kindly provided by Dr.~Rolf Rabenseifner at HLRS, in connection with a course on PGAS programming~\cite{PGAS_course2015}} that was tested in previous work \cite{lagraviere2019performance}. The UPC++ implementation is derived from this UPC implementation. Both the UPC and UPC++ codes implement a 2D heat equation solver using halo-exchange with a single layer of ghost cells to communicate data between threads.
We use $X$ and $Y$ to denote the number of cells in each dimension of the domain. Each processing elements receives a rectangular sub-domain of approximately equal size. In order to ensure the communication of data to adjacent subdomain(s) we use the well-known halo exchange technique \cite{davies1970domain}. 
In both our implementations of the 2D Heat Equation the domains receive their initial values before the computation starts. The heat diffusion is then computed by performing $N$ time steps. For each time step, each processing-element performs two tasks: computing the diffusion by updating each cell in its subdomain, and communicating the newly computed results on the domain boundary to adjacent subdomains. All time steps perform the same amount of computation, but due to cache effects and the unpredictability of network communication their execution time can vary. For benchmarking purposes it is important to obtain the average time per time step. Typically, running for $N = 1000$ time steps is sufficient to do so.

\subsubsection{Sparse matrix-vector multiplication}
\label{spmvKernel}
A general matrix-vector multiplication is compactly and mathematically defined by the following formula: $y=M x$.
In this study, without loss of generality, we assume that the matrix $M$ is square, having $n$ rows and $n$ columns. 
Both the input vector and the result vector, respectively denoted by $x$ and $y$, are of length $n$.
Then, to compute element number $i$ of the result vector $y$ we apply the following general formula (using zero-based indices):
\begin{equation}
y(i) = \sum_{0\le j< n} M(i,j)x(j).
\label{eqn:MV}
\end{equation}

$M$ is called a {\em sparse}
matrix, if most of the $M(i,j)$ values are zero.
In this case the above formula becomes unnecessarily expensive from a computational point of view.
A more economic formula for computing $y(i)$ in a sparse matrix-vector
multiplication (SpMV) is thus:
\begin{equation}
y(i) = \sum_{M(i,j)\neq 0} M(i,j)x(j),
\label{eqn:SpMV}
\end{equation}
This formula  takes into account only the nonzero values of matrix $M$ on each row.
Because only the nonzero values are used, it is then memory-wise unnecessarily expensive to store all
the $n^2$ values of a sparse matrix.
As a result various compact storage formats for sparse matrices have been adopted, such as:
\begin{itemize}
    \item the coordinate format (COO);
    \item the compressed sparse row format (CSR);
    \item the compressed sparse column format (CSC);
    \item and the EllPack format~\cite{Grimes1979}.
\end{itemize} 

For sparse matrices that have a homogeneous number of nonzero values per row, it is usual to use the EllPack storage format. The EllPack storage format usually uses two 2D tables.
These two 2D tables are of the same size: having $n$ rows and
the number of columns equaling the maximum number of non-zeros per row.
The first 2D table contains all the nonzero values of the sparse matrix.
Whereas the second 2D  table contains the corresponding column indices of
the non-zeros. 
Explicitly stored zeros are used for padding when the rows that have fewer than the maximum number of non-zeros.
Additionally, if we assume that all the values on the
main diagonal of a sparse matrix $M$ are nonzero, which is applicable to
most scientific applications, it is beneficial to split $M$ so that:
\begin{equation}
M=D+A \label{eqn:SpMV2}
\end{equation}
In this formula, $D$ is the main diagonal of $M$, while $A$ contains the off-diagonal part of $M$.
Then, we can use a modified EllPack format where the main diagonal $D$ is stored as a 1D array of length $n$.
There is no need to store the column indices of these nonzero
diagonal values, because their column indices equal the row indices by
definition. 
We suppose that $r_{nz}$ now denotes the maximum number of non-zero
off-diagonal values per row.
Thus, for storing the values in the off-diagonal part $A$, it is
usual to use two 1D arrays both of length $n\cdot r_{nz}$ (instead of two $n\times r_{nz}$ 2D tables).
Therefore, one 1D array contains the
off-diagonal values consecutively row by row, whereas the other 1D array contains the corresponding integer column indices \cite{lagraviere2019performance}.

The goal of this chapter is to present a detailed view of the our implementations using UPC++. In addition, for comparison purposes, we also show the corresponding UPC code implementing the same kernel as in UPC++. We want to emphasize the differences in the UPC and UPC++ programming models and their consequences in terms of code.

\subsection{Definitions}
In our UPC++ implementations of SpMV and the Heat Equation we use repeatedly \emph{global pointers} and communication function such as \emph{rget} or \emph{rput}. In this section we present definitions for these terms and functionalities.

\subsubsection{UPC++'s shared arrays and global pointers: PGAS in action}
A UPC++ program can allocate global memory in shared segments, which are accessible by all processes. A
global pointer points at storage within the global memory, and is declared as follows:

{\tt upcxx::global\_ptr<int> gptr = upcxx::new\_<int>(upcxx::rank\_me());}

The call to {\tt upcxx::new\_<int>} allocates a new integer in the calling process’s shared segment, and returns a
global pointer ({\tt upcxx::global\_ptr}) to the allocated memory. Each process has its own private pointer (gptr) to an integer in its local shared segment. By contrast, a conventional C++ dynamic allocation (int *mine = new int) will be in private local memory. Note
that we use the integer type in this paragraph as an example, but any type T can be allocated using the
{\tt upcxx::new\_<T>()} function call \cite{upc++programmersGuide}.

In our implementation of both SpMV and the Heat Equation we use shared arrays and global pointers. The declaration and allocation of shared arrays is a multiple step process, as described in Listing \ref{code-UPC++-v3-prep} (page \pageref{code-UPC++-v3-prep}) with a buffer called {\tt share\_receive\_buffer}.

\subsubsection{UPC++: rget and rput}
Both UPC++ SpMV codes, i.e.~SpMV using block-wise data transfer and SpMV using message condensing and consolidation use the UPC++ function {\tt upcxx::rput}. This function is used to send data from one thread to another asynchronously by default. In our implementation of the UPC++ Heat Equation 2D we use the UPC++ function {\tt upcxx::rget}.
Both functions are part of the one-sided communication model that is implemented in UPC++.  These operations initiate transfer of the value object to (put) or from
(get) the remote process; no coordination is needed with the remote process since it is a one-sided communication. 
Like many asynchronous communication operations, rget and rput default to returning a future object \cite{Futures_and_promises} that becomes ready when the transfer is complete \cite{upc++programmersGuide}.

\subsection{Implementations of sparse matrix-vector multiplication}
\label{spmvImplementation}
In our study we have implemented multiple versions of the Sparse Matrix-Vector Multiplication (SpMV) both in UPC and UPC++. The difference between these versions lies essentially in the communication methodology. Thus, in this section we describe 4 implementations of SpMV using a modified ELLPack format, as described in Section \ref{spmvKernel}.

These 4 versions are:
\begin{itemize}
    \item UPC SpMV using Block-wise data transfer: Section \ref{blockWiseImplementation}, Listing \ref{code-UPC-v2}, based on our previous study \cite{lagraviere2019performance};
    \item UPC++ SpMV using Block-wise data transfer: Section \ref{blockWiseImplementation}, Listing \ref{code-UPC++v2};
    \item UPC SpMV using message condensing and consolidation: Section \ref{cleverComms}, Listing \ref{code-UPC-v3},  based on our previous study \cite{lagraviere2019performance};
    \item UPC++ SpMV using message condensing and consolidation: Section\ref{cleverComms}, Listings \ref{code-UPC++-v3}, \ref{code-UPC++-v3-prep}.
\end{itemize}

Note that in UPC {\tt upcxx::rank\_me} corresponds to {\tt MYTHREAD}, meaning 'id number of calling thread' and {\tt upcxx::rank\_n} corresponds to THREADS, meaning 'total number of threads used for the current run'.

For both UPC and UPC++ the code can be summarized as follows: first, we read and distribute the data in memory, then we prepare the data (particularly preparing data exchange buffers), and after that the actual SpMV computation occurs.

\subsubsection{Explicit thread privatization}
In both UPC and UPC++ codes presented in the following sections, we have used as much as possible a strategy that we call \textit{explicit thread privatization}. The goal of this technique is to ensure that each thread accesses (read and write) its own local data. Doing so means that no implicit communication is triggered and no-illegal memory access is performed, thus accessing the data thanks to \textit{explicit thread privatization} delivers the best possible performance. To achieve such a goal, we use the well-known technique of casting \textit{pointers-to-shared} to \textit{pointers-to-local}. To ensure that we create pointers that point to data with affinity to the calling thread ({\tt MYTHREAD} in UPC, {\tt upcxx:rank\_me} in UPC++), we use the {\tt BLOCKSIZE} to point to the local data in the computation loop. This technique is illustrated in the  Listing \ref{upcETP} for UPC and in Listing \ref{upcxxETP} for UPC++. 

\begin{lstlisting}[caption=Explicit thread privatization in UPC (Complete code in Listing \ref{code-UPC-v2}),label=upcETP]
for (int mb=0; mb<mythread_nblks; mb++) {
  int offset = (mb*THREADS+MYTHREAD)*BLOCKSIZE;
  /* casting shared pointers to local pointers */
  double *loc_y = (double*) (y+offset);
  //...
}
\end{lstlisting}
\begin{lstlisting}[caption=Explicit thread privatization in UPC++,label=upcxxETP]
for (mb=0; mb<mythread_nblks; mb++) {
  offset = (mb*upcxx::rank_n()S+upcxx::rank_me())*BLOCKSIZE;
  /* casting shared pointers to local pointers */
  double *loc_y = (double*) (y[upcxx::rank_me()]+offset);
  //...
}
\end{lstlisting}


\subsubsection{Block-Wise Data Transfer between Threads}
\label{blockWiseImplementation}
In this section, we present a summarized view of UPC and UPC++ SpMV implementing block-wise data transfer, as seen in Listings \ref{code-UPC-v2} and \ref{code-UPC++v2}.
The UPC SpMV using block-wise data transfer is the same as the one we used in our previous study \cite{lagraviere2019performance}. The UPC++ SpMV using block-wise data transfer is derived from this aforementioned UPC version, in the sense that it implements the same logic of computation and communication. To the exception that in UPC we used {\tt upc\_memget}, which gets data from another thread's shared memory , for communication and for UPC++ we used {\tt upcxx::rput}, which sends data to another thread's shared memory. 

In these versions of UPC and UPC++ SpMV, the communication of data is done block-wise. This means that for UPC if {\tt Thread A} requires one element {\tt X}  to {\tt Thread B}, a block of data of size {\tt BLOCKSIZE} containing element {\tt X} will be sent from {\tt Thread B}  to {\tt Thread A}. In other words, each needed block is transported in its {\em entirety}, independent of the actual number of values needed in that block.
For UPC++ the operation is initiated from the sending thread as we have used a one-sided {\tt put} function whereas in UPC we have used a one-sided {\tt get} function.

\begin{lstlisting}[language=C,caption=UPC implementation
  of SpMV by block-wise communication \cite{lagraviere2019performance},
label=code-UPC-v2]
//Reading the data set and data distribution over THREADS then prepare the data for computation
double *mythread_x_copy = (double*) malloc(n*sizeof(double));
/* Prep-work: check for each block of x whether it has values needed by MYTHREAD; make a private boolean array 'block_is_needed' of length nblks */
// ....
/* Transport the needed blocks of x into mythread_x_copy */
for (int b=0; b<nblks; b++)
   if (block_is_needed[b])
     upc_memget(&mythread_x_copy[b*BLOCKSIZE],&x[b*BLOCKSIZE],
                min(BLOCKSIZE,n-b*BLOCKSIZE)*sizeof(double));
/* SpMV: each thread only goes through its designated blocks */
int mythread_nblks = nblks/THREADS+(MYTHREAD<(nblks%THREADS)?1:0);
for (int mb=0; mb<mythread_nblks; mb++) {
  int offset = (mb*THREADS+MYTHREAD)*BLOCKSIZE;
  /* casting shared pointers to local pointers */
  double *loc_y = (double*) (y+offset);double *loc_D = (double*) (D+offset);
  double *loc_A = (double*) (A+offset*r_nz);int *loc_J = (int*) (J+offset*r_nz);
  /* computation per block */
  for (int k=0; k<min(BLOCKSIZE,n-offset); k++) {
    double tmp = 0.0;
    for (int j=0; j<r_nz; j++)
      tmp += loc_A[k*r_nz+j]*mythread_x_copy[loc_J[k*r_nz+j]];
    loc_y[k] = loc_D[k]*mythread_x_copy[offset+k] + tmp;
  }
}
\end{lstlisting}

\begin{lstlisting}[language=C,caption=UPC++ SpMV implementation of 'Block-Wise Data Transfer between Threads', label=code-UPC++v2]
//Reading the data set and data distribution over THREADS then prepare the data for computation
double *mythread_x_copy = (double*) malloc(n*sizeof(double));
/* Prep-work: check for each block of x whether it has values needed by upcxx::rank_me(); make a private boolean array 'block_is_needed' of length nblks */
//Creating a future in order to store all communication requests
upcxx::future<> futureComms = upcxx::make_future();
/* SpMV: each thread only goes through its designated blocks */ 
int mythread_nblks = nblks/upcxx::rank_n()+(upcxx::rank_me()<(nblks%upcxx::rank_n())?1:0);
for (int mb=0; mb<mythread_nblks; mb++) {
  int offset = (mb*upcxx::rank_n()+upcxx::rank_me())*BLOCKSIZE;
  /* casting shared pointers to local pointers */
  double *loc_y = (double*) (y[upcxx::rank_me()]+offset); double *loc_D = (double*) (D[upcxx::rank_me()]+offset);
  double *loc_A = (double*) (A[upcxx::rank_me()]+offset*r_nz);
  int * loc_J = (int*)I[upcxx::rank_me()].local()+offset*r_nz;
  /* computation per block */
  for (int k=0; k<min(BLOCKSIZE,n-offset); k++) {
    double tmp = 0.0;
    for (int j=0; j<r_nz; j++)
      tmp += loc_A[k*r_nz+j]*mythread_x_copy[loc_J[k*r_nz+j]];
    loc_y[k] = loc_D[k]*mythread_x_copy[offset+k] + tmp;
  }
/* Transport the needed blocks of x into mythread_x_copy */
for (int targetThread=0;targetThreads<numberOfTargets, targetThread++)
        for (int b=0; b<nblks; b++)
            if (block_is_needed[b])
                futureComms = upcxx::when_all(futureComm, upcxx::rput(&mythread_x_copy[b*BLOCKSIZE], 
		         x[targetThread]+(b*BLOCKSIZE), min(BLOCKSIZE,n-b*BLOCKSIZE)));
//Waiting for all communication to be completed
futureComms.wait();
\end{lstlisting}
 
In the UPC++ implementation (Listing \ref{code-UPC++v2}) we use a \emph{future} \cite{Futures_and_promises}, called {\tt futureComms} in conjunction with {\tt upcxx::when\_all}. 
Unlike standard C++ futures, UPC++ futures and promises
are used to manage asynchronous dependencies within a
thread and not for direct communication between threads or
processes. A future thus represents the consumer side of a
non-blocking operation. Each non-blocking operation has an
associated promise object, which is created either explicitly by
the user or implicitly by the runtime when the non-blocking
operation is invoked. A promise represents the producer side
of the operation, and it is through the promise that the results
of the operation are supplied and its dependencies fulfilled. A
user can pass a promise to a UPC++ communication operation,
which registers a dependency on the promise and subsequently
fulfills the dependency when the operation completes. The
same promise can be passed to multiple communication operations, and through a single wait call on the associated
future, the user can be notified when all the operations have
completed \cite{bachan2019upc++}.

The future conjoining begins by invoking  {\tt upcxx::make\_future} to construct a trivially ready future  {\tt futureComms}.
A new future, returned by {\tt upcxx::rput} is passed to the  {\tt upcxx::when\_all} function, in combination with the previous future, {\tt futureComms}. The  {\tt upcxx::when\_all} constructs a new future
representing readiness of all its arguments, and returns a future with a
concatenated results tuple of the arguments. By setting {\tt futureComms} to the future returned by {\tt upcxx::when\_all}, we
can extend the conjoined futures. Once all the processes are linked into the conjoined futures, we simply
wait on the final future, i.e.~the {\tt futureComms.wait} call \cite{upc++programmersGuide}.

In the UPC++ implementation (Listing \ref{code-UPC++v2}, \pageref{code-UPC++v2}) we have used Futures in addition to {\tt upcxx::when\_all}, {\tt upcxx::rput} and {\tt <upcxx::future>.wait}, which implement one-sided, asynchronous, bulk communication. And in UPC we have used, one-sided, bulk communication using {\tt upc\_memget}. By default, all communications in UPC++ are asynchronous, which, in terms of code corresponds to use of at least one instruction for sending or receiving data and one instruction to wait for the completion of the communication. In our UPC++ implementation, we have chosen to distinguish the sending of data performed by {\tt upcxx::rput} and the waiting for completion performed by {\tt futureComms.wait}. This means that multiple communications are launched in a non-blocking manner without having to wait for one-another in order to start transmitting data. This is done at the expense and with a measured risk of bandwidth (between CPU and RAM) and network (The test system in our experiments uses \textit{Infiniband}, as described in section \ref{hardware}).


\subsubsection{Message condensing and consolidation}
\label{cleverComms}
In this versions of UPC and UPC++ SpMV, we implement a different communication technique. Contrary to the block-wise data transfer presented in section \ref{blockWiseImplementation}, where each communication request is of a full block size ({\tt BLOCKSIZE}), here, we employ a communication strategy where each thread communicates the strict amount of required data to other threads. In other words, the length of a message from thread {\tt A}
to thread {\tt B} equals the number of {\em unique} values in the {\tt k}
blocks owned by {\tt A} that are needed by {\tt B}, times the number of bytes required to represent each value (8 byte double in our experiments).
All the between-thread messages are thus condensed and consolidated.

Listing \ref{code-UPC++-v3-prep} (page \pageref{code-UPC++-v3-prep})  and Listing  \ref{code-UPC++-v3} (page \pageref{code-UPC++-v3}) present respectively the preparation step and the computation step of the UPC++ version of SpMV using message condensing and consolidation.
Listing \ref{code-UPC-v3} (page \pageref{code-UPC-v3}) presents the preparation and computation steps of the UPC version of SpMV using message condensing and consolidation.

This communication patterns and the use of the UPC++ function {\tt upcxx::rput} leads to the necessity of a \emph{Shared Receive Buffer} (SRB). This SRB will be accessed by each thread that needs to communicate data, and then the receiving thread will get the data from this buffer after the all threads are done communicating. 
In both UPC and UPC++ implementations we have the necessity for the SRB. In UPC++ the implementation of the SRB is different and slightly more complicated than in UPC. SRBs in UPC++ require the use of a two-dimensional vector and the corresponding declarations and function calls to populate it and broadcast the information across all threads ({\tt upcxx::broadcast}).

\begin{lstlisting}[language=C,caption=An improved UPC implementation
  of SpMV by message condensing and consolidation \cite{lagraviere2019performance},
label=code-UPC-v3]
/* Allocation of the five shared arrays x, y, D, A, J */
// ...
/* Allocation of an additional private x array per thread */
double *mythread_x_copy = (double*) malloc(n*sizeof(double));

/* Preparation step: create and fill the thread-private arrays of int *mythread_num_send_values, int *mythread_num_recv_values, int **mythread_send_value_list, int **mythread_recv_value_list, double **mythread_send_buffers. Also, shared_recv_buffers is prepared. */
// ....

/* Communication procedure starts */
int T,k,mb,offset;
double *local_x_ptr = (double *)(x+MYTHREAD*BLOCKSIZE);
for (T=0; T<THREADS; T++)
  if (mythread_num_send_values[T]>0) /* pack outgoing messages */
    for (k=0; k<mythread_num_send_values[T]; k++)
      mythread_send_buffers[T][k] = local_x_ptr[mythread_send_value_list[T][k]];

for (T=0; T<THREADS; T++)
  if (mythread_num_send_values[T]>0) /* send out messages */
    upc_memput(shared_recv_buffers[T*THREADS+MYTHREAD], mythread_send_buffers[T],
               mythread_num_send_values[T]*sizeof(double));

upc_barrier;

int mythread_nblks = nblks/THREADS+(MYTHREAD<(nblks%THREADS)?1:0);
for (mb=0; mb<mythread_nblks; mb++) {/* copy own x-blocks */
  offset = (mb*THREADS+MYTHREAD)*BLOCKSIZE;
  memcpy(&mythread_x_copy[offset], (double *)(x+offset), min(BLOCKSIZE,n-offset)*sizeof(double));
}

for (T=0; T<THREADS; T++)
  if (mythread_num_recv_values[T]>0) {/* unpack incoming messages */
    double *local_buffer_ptr = (double *)shared_recv_buffers[MYTHREAD*THREADS+T];
    for (k=0; k<mythread_num_recv_values[T]; k++)
      mythread_x_copy[mythread_recv_value_list[T][k]] = local_buffer_ptr[k];
  }
/* Communication procedure ends */

/* SpMV: each thread only goes through its designated blocks */
for (mb=0; mb<mythread_nblks; mb++) {
  offset = (mb*THREADS+MYTHREAD)*BLOCKSIZE;
  /* casting shared pointers to local pointers */
  double *loc_y = (double*) (y+offset);
  double *loc_D = (double*) (D+offset);
  double *loc_A = (double*) (A+offset*r_nz);
  int *loc_J = (int*) (J+offset*r_nz);
  /* computation per block */
  for (k=0; k<min(BLOCKSIZE,n-offset); k++) {
    double tmp = 0.0;
    for (int j=0; j<r_nz; j++)
      tmp += loc_A[k*r_nz+j]*mythread_x_copy[loc_J[k*r_nz+j]];
    loc_y[k] = loc_D[k]*mythread_x_copy[offset+k] + tmp;
  }
}
\end{lstlisting}

\begin{lstlisting}[language=C,caption=An improved UPC++ implementation
  of SpMV by message condensing and consolidation: Preparation step,
label=code-UPC++-v3-prep]
/* Allocation of the five shared arrays x, y, D, A, J */
// Preparation Begins
/* Allocation of an additional private x array per thread */
double *mythread_x_copy = (double*) malloc(n*sizeof(double));

/* Preparation step: create and fill the thread-private arrays of int *mythread_num_send_values, int *mythread_num_recv_values, int **mythread_send_value_list, int **mythread_recv_value_list, double **mythread_send_buffers.*/


//Preparation of the shared_receive_buffer
vector<std::vector<global_ptr<double> > >shared_receive_buffer(upcxx::rank_n());
std::vector<global_ptr<double>> temp(upcxx::rank_n());
for(int i = 0; i < upcxx::rank_n(); i++)
	temp[i] = upcxx::new_array<double>(localReceiveCount[i]);

shared_receive_buffer[upcxx::rank_me()] = temp;

for(int i = 0; i<upcxx::rank_n();i++)
	shared_receive_buffer[i] = upcxx::broadcast(shared_receive_buffer[i], i).wait();
//Creating a "empty" future, used to store communication requests
upcxx::future<> futureComms = upcxx::make_future();
\end{lstlisting}

\begin{lstlisting}[language=C,caption=An improved UPC++ implementation
  of SpMV by message condensing and consolidation: Computation step,
label=code-UPC++-v3]
//computation starts
/* SpMV: each thread only goes through its designated blocks */
for (mb=0; mb<mythread_nblks; mb++) {
  offset = (mb*upcxx::rank_n()S+upcxx::rank_me())*BLOCKSIZE;
  /* casting shared pointers to local pointers */
  double *loc_y = (double*) (y[upcxx::rank_me()]+offset);
  double *loc_D = (double*) (D[upcxx::rank_me()]+offset);
  double *loc_A = (double*) (A[upcxx::rank_me()]+offset*r_nz);
  int * loc_J = (int*)I[upcxx::rank_me()].local()+offset*r_nz;
  /* computation per block */
  for (k=0; k<min(BLOCKSIZE,n-offset); k++) {
    double tmp = 0.0;
    for (int j=0; j<r_nz; j++)
      tmp += loc_A[k*r_nz+j]*mythread_x_copy[loc_J[k*r_nz+j]];
    loc_y[k] = loc_D[k]*mythread_x_copy[offset+k] + tmp;
  }
}

/* Communication procedure starts */
int T,k,mb,offset;
double *local_x_ptr = (double *)(x+upcxx::rank_me()*BLOCKSIZE);
for (T=0; T<upcxx::rank_n()S; T++)
  if (mythread_num_send_values[T]>0) /* pack outgoing messages */
    for (k=0; k<mythread_num_send_values[T]; k++)
      mythreadsend_buffers[T][k] = local_x_ptr[mythreadsend_value_list[T][k]];

for (T=0; T<upcxx::rank_n()S; T++) {
  if (mythread_num_send_values[T]>0) /* send out messages */
	futureComms.when_all(futureComms, upcxx::rput( mythread_send_buffers[T], shared_recv_buffers[T][upcxx::rank_me()], mythread_num_send_values[T]));
}
//waiting for all communication to be completed
futureComms.wait();
upc_barrier;
//Receiving data, unpacking it into mythread_x_copy
for (T=0; T<upcxx::rank_n()S; T++)
  if (mythreadnum_recv_values[T]>0) {/* unpack incoming messages */
    double *local_buffer_ptr = (double *)shared_recv_buffers[upcxx::rank_me()*upcxx::rank_n()S+T];
    for (k=0; k<mythread_num_recv_values[T]; k++)
      mythread_x_copy[mythread_recv_value_list[T][k]] = local_buffer_ptr[T][k];
  }
/* Communication procedure ends */
\end{lstlisting}

\subsection{Implementations of heat equation 2D}
\label{heatEquationImplementation}
The UPC code used in this section is identical to the one used in our previous study \cite{lagraviere2019performance}. The UPC code was kindly provided by Dr.~Rolf Rabenseifner at HLRS, in connection with a short course on PGAS programming~\cite{PGAS_course2015}.
The UPC++ code for Heat Equation 2D is inspired from the UPC code in that it uses identical policies for data distribution, computation and data communication through halo data exchange.
Thus, in this section we propose the following implementations of a solver for the Heat Equation on a 2-dimensional domain:
\begin{itemize}
    \item UPC implementation is presented in the following listings:
        \begin{itemize}
        \item "Scratch" arrays for data exchange in Listing \ref{scratchUPC} (page \pageref{scratchUPC})
        \item Halo data exchange in Listing \ref{code-2D-comm} (page \pageref{code-2D-comm})
        \end{itemize}
    \item UPC++ implementation is presented in the following listings:
        \begin{itemize}
        \item "Scratch" arrays for data exchange in Listing \ref{scratchUPC++} (page \pageref{scratchUPC++})
        \item Halo data exchange in Listing \ref{upc++-2D-comm} (page \pageref{upc++-2D-comm})
        \end{itemize}
\end{itemize}

In our implementations of UPC and UPC++ Heat Equation, the global 2D solution domain is rectangular, so the UPC and UPC++ threads are arranged as a 2D processing grid, with {\tt mprocs} rows and {\tt nprocs} columns. (Note {\tt upcxx::rank\_n} equals {\tt mprocs*nprocs}.)
Each thread is thus identified by an index pair
{\tt (iproc,kproc)}, where
{\tt iproc = upcxx::rank\_me / nprocs} and {\tt kproc = upcxx::rank\_me \% nprocs}.
The global 2D domain, of dimension $M\times N$, is evenly divided
among the threads. Each thread is responsible for a 2D sub-domain of
 dimension $m\times n$, which includes a surrounding halo layer needed for communication with the neighboring threads. Reminder: In UPC++ {\tt upcxx::rank\_me} corresponds to {\tt MYTHREAD} in UPC and, in UPC++, {\tt upcxx::rank\_n} corresponds to {\tt THREADS}, in UPC.

 In both UPC and UPC++ implementations we use a halo data exchange, which is needed between the neighboring threads. The halo data exchange performs communication, where each thread calls {\tt upc\_memget} for UPC or {\tt upcxx::rget} for UPC++, on each of the four sides of its subdomain (if a neighboring thread exists). In the vertical direction, the values to be transferred from the upper and lower neighbors already lie contiguously in the memory of the owner threads. There is thus no need to explicitly pack the messages. In the horizontal direction, however, message packing is needed before {\tt upc\_memget} can be invoked towards the left and right neighbors.
 
Listing \ref{scratchUPC} presents additional data structure which is needed with packing and unpacking the horizontal messages. Listing \ref{scratchUPC++}, presents the same idea implemented in UPC++. This short example of code also shows the additional instructions required by the use of UPC++: the declaration of shared arrays and their allocation in memory requires more code than in UPC.

\begin{lstlisting}[language=C,caption=Scratch arrays for UPC halo exchange of non-contiguous data \cite{lagraviere2019performance},label=scratchUPC]
shared [] double * shared xphivec_coord1first[THREADS];
shared [] double * shared xphivec_coord1last [THREADS];
double *halovec_coord1first, *halovec_coord1last;

xphivec_coord1first[MYTHREAD] =
  (shared [] double *) upc_alloc((m-2)*sizeof(double)); 
xphivec_coord1last[MYTHREAD]  =
  (shared [] double *) upc_alloc((m-2)*sizeof(double));
halovec_coord1first = (double *) malloc((m-2)*sizeof(double)); 
halovec_coord1last  = (double *) malloc((m-2)*sizeof(double));
\end{lstlisting}
 
\begin{lstlisting}[language=C,caption=Scratch arrays for UPC++ halo exchange of non-contiguous data,label=scratchUPC++]
std::vector<upcxx::global_ptr<double>> xphivec_coord1first;
std::vector<upcxx::global_ptr<double>> xphivec_coord1last;
xphivec_coord1first.resize(upcxx::rank_n());
xphivec_coord1last.resize(upcxx::rank_n());
xphivec_coord1first[upcxx::rank_me()] = upcxx::new_array<double>(m-2);
xphivec_coord1last[upcxx::rank_me()] = upcxx::new_array<double>(m-2);
for (int i = 0; i < upcxx::rank_n(); i++) {
	xphivec_coord1first[i] = upcxx::broadcast(xphivec_coord1first[i], i).wait();
	xphivec_coord1last[i] = upcxx::broadcast(xphivec_coord1last[i], i).wait();
}

phivec_coord1first = (double *) xphivec_coord1first[upcxx::rank_me()].local(); 
phivec_coord1last  = (double *) xphivec_coord1last[upcxx::rank_me()].local();

double *halovec_coord1first = (double *) calloc((m-2),	sizeof(double)); 
double *halovec_coord1last  = (double *) calloc((m-2),  sizeof(double));
\end{lstlisting}
 
The communication implementation through halo exchange is different in UPC and UPC++. Listing \ref{code-2D-comm} presents the communication performed in UPC using {\tt upc\_memget}, and performing a packing of the horizontal data beforehand. Listing \ref{upc++-2D-comm} presents the implementation of halo data exchange using {\tt upcxx::rget} and performing a packing of the horizontal data beforehand.
 
 In the UPC++ version it is important to notice that we store futures {\tt upcxx::future} in a {\tt std::vector} called {\tt getRequests}, this vector is then used in the function {\tt waitForGetRequestsToEnd}. This function calls the UPC++ {\tt wait} function on each {\tt upcxx::future} contained in the vector {\tt getRequests}. This is different technique to wait for all communication to complete compared to the one used in Listing \ref{code-UPC++-v3} (page \pageref{code-UPC++-v3}), where we used the combination of {\tt upcxx::when\_all} and {\tt upcxx::wait}.
 
 The communication technique used in UPC and UPC++ relies on the same halo data exchange model, however, the amount of code needed in UPC++ is greater than the one needed in UPC. As explained earlier in this paper, this difference in the amount of code is directly related to the way shared arrays are declared in UPC and UPC++ and the fact that communication in UPC++ is done in two steps: {\tt upcxx::rget} and the {\tt wait} statement whereas in UPC a single function call to {\tt upc\_memget} is needed.

\begin{lstlisting}[language=C,caption=The halo data exchange function of an existing UPC 2D heat equation solver \cite{lagraviere2019performance},label=code-2D-comm]
#define idx(i,k) ((i)*n+(k)) 
#define rank(ip,kp) ((ip)*nprocs+(kp)) 
void halo_exchange_intrinsic() {
  double* phi = (double *) xphi[MYTHREAD]; int i,k;
  /* packing messages for the horizontal direction */
  if (kproc>0) {
    double *phivec_coord1first = (double *) xphivec_coord1first[MYTHREAD];
    for (i=0; i<m-2; i++)       { phivec_coord1first[i] = phi[idx(i+1,1)]; }
  }
  if (kproc<nprocs-1) {
    double *phivec_coord1last = (double *) xphivec_coord1last[MYTHREAD];
    for (i=0; i<m-2; i++)       { phivec_coord1last[i] = phi[idx(i+1,n-2)]; }
  }
  upc_barrier;
  /* message transfer and unpacking (needed for the horizontal direction) */
  if (kproc>0) {
    upc_memget(halovec_coord1first, xphivec_coord1last[rank(iproc,kproc-1)], (m-2)*sizeof(double));
    for (i=1; i<m-1; i++)       { phi[idx(i,0)] = halovec_coord1first[i-1]; }
  }
  if (kproc<nprocs-1) {
    upc_memget(halovec_coord1last, xphivec_coord1first[rank(iproc,kproc+1)], (m-2)*sizeof(double));
    for (i=1; i<m-1; i++)       { phi[idx(i,n-1)] = halovec_coord1last[i-1]; }
  }
  if (iproc>0)
    upc_memget(&phi[idx(0,1)], &(xphi[rank(iproc-1,kproc)][idx(m-2,1)]), (n-2)*sizeof(double));
  if (iproc<mprocs-1)
    upc_memget(&phi[idx(m-1,1)], &(xphi[rank(iproc+1,kproc)][idx(1,1)]), (n-2)*sizeof(double));
}
\end{lstlisting}
 
  
\begin{lstlisting}[language=C,caption=The halo data exchange function of our  UPC++ implementation for 2D heat equation solver ,label=upc++-2D-comm]
//Standard CPP vector containing UPC++ Futures
std::vector<upcxx::future<>> getRequests(4)
void halo_exchange_intrinsic() {
	int i,k;             
	/* packing messages for the horizontal direction */
	if (kproc>0)
		for (i=0;i<m-2;i++) { phivec_coord1first[i] = phi[idx(i+1,1)]; }
			
	if (kproc<nprocs-1)
		for (i=0;i<m-2;i++) { phivec_coord1last[i] = phi[idx(i+1,n-2)]; }
		
	upcxx::barrier();
	/* message transfer and unpacking (needed for the horizontal direction) */
	if (iproc>0) {
		getRequests[0] = upcxx::rget(xphi[getRank(iproc-1,kproc)]+idx(m-2,1), &phi[idx(0,1)], n-2); }
	if (iproc<mprocs-1) {
		getRequests[1] = upcxx::rget(xphi[getRank(iproc+1,kproc)]+idx(1,1), &phi[idx(m-1,1)], n-2); }
	if (kproc>0) {
		getRequests[2] = upcxx::rget(xphivec_coord1last[getRank(iproc,kproc-1)], halovec_coord1first, m-2);
		for (i=1;i<m-1;i++)	{phi[idx(i,0)] = halovec_coord1first[i-1];}}
		
	if (kproc<nprocs-1) {
		getRequests[3] = upcxx::rget(xphivec_coord1first[getRank(iproc,kproc+1)], halovec_coord1last, m-2);
		for (i=1;i<m-1;i++) { phi[idx(i,n-1)] = halovec_coord1last[i-1];}}
	waitForGetRequestsToEnd();
}
void waitForGetRequestsToEnd() {
	if (iproc>0)            { getRequests[0].wait(); }
	if (iproc<mprocs-1)     { getRequests[1].wait(); }
	if (kproc>0)            { getRequests[2].wait(); }
	if (kproc<nprocs-1)     { getRequests[3].wait(); }
}
\end{lstlisting}

\section{Experimental Setup}
\label{experimentalSetup}
\label{software}
\label{hardware}

Our performance measurements focus on many-core and multi-node hardware architectures. 

For the many-core experiments we run our UPC and UPC++ codes on a machine equipped with one Intel Xeon Phi 7250 (Knights Landing, or KNL) processor, which is equipped with 16GB of high speed MCDRAM. Due to the hardware structure of the KNL processor we can select the way MCDRAM is seen (addressed and accessed) by the operating system and the programs. For our experiments we chose the \emph{flat mode}. When the Knights Landing processor is booted in \emph{flat mode}, the entirety of the MCDRAM is used as addressable memory. MCDRAM as addressable memory shares the physical address space with DDR4, and is also cached by the L2 cache. With respect to Non Uniform Memory Access (NUMA) architecture, the MCDRAM portion of the addressable memory is exposed as a separate NUMA node without cores, with another NUMA node containing the DDR4 memory \cite{intelKnlMemoryStuff}. We use \emph{numactl} to explicitly place data in the MCDRAM NUMA node. KNL processors can alternatively be booted in \emph{cache mode}, which uses 
the MCDRAM as cache which is thus transparent for the operating system, and in \emph{hybrid mode} which used 8GB as cache and 8GB as addressable memory.
In previous work we found that for problems which fit entirely within the 16GB of MCDRAM, which is the case for our test instances, the performance of \emph{cache mode} is only marginally lower than that of \emph{flat mode} \cite{langguth2017porting}. 
On this machine we used Intel Compiler version 17.0.0 to compile UPC and UPC++ compilers and runtimes environments, as well as our programs.

The Abel computer cluster~\cite{abel_cluster} was used to run all the
UPC and UPC++ codes on multi-node and measure their time usage. Each compute node on Abel
is equipped with two Intel Xeon E5-2670 2.6~GHz 8-core CPUs and
 64~GB of RAM. The interconnect between the nodes is FDR InfiniBand 
(56~Gbits/s). In our previous study \cite{lagraviere2019performance}, we measured the  memory bandwidth per node on Abel and obtained 75 GB/s; we also measured the inter-node communication bandwidth and obtained about 6 GB/s. On this supercomputer we have used nodes providing access to 16 physical cores, and for all our runs we always have used the maximum amount of physical cores per node. In the following, when speaking of cores we mean \emph{physical core}. We have not used any kind of purely logical cores offered  by Intel's HyperThreading technology.

The Berkeley UPC~\cite{Berkeley_UPC} version 2.24.2 was used for
compiling and running all our UPC implementations of SpMV and Heat Equation. The compilation procedure
involved first a behind-the-scene translation from UPC to C done remotely
 at Berkeley via HTTP, with the translated C code being then compiled
locally on Abel using Intel's {\tt icc} compiler version
15.0.1. The compilation options are {\tt -O3 -std=gnu99}.

The Berkeley UPC++~\cite{upc++OfficialWebpage} version 1.0 (2019.9.0) was used for compiling and running all our UPC++ implementation of SpMV and Heat Equation. UPC++ relies on a local compiler and MPI installation (i.e.~located on the supercomputer). MPI is used to spawn UPC++ process on each node. GNU G++ version 7.2.0 was used to compile C++ code, and OpenMPI version 3.1.2 was used for process spawning.

Both UPC and UPC++ use GASNet(-EX) as an under-layer to ensure communication between threads, processes or cores and nodes. This also means, that UPC and UPC++ are APIs for GASNet. GASNet is a language-independent networking middleware layer that provides network-independent, high-performance communication primitives including Remote Memory Access (RMA) and Active Messages (AM). It has been used to implement parallel programming models and libraries such as UPC, UPC++, Co-Array Fortran, Legion, Chapel, and many others. The interface is primarily intended as a compilation target and for use by runtime library writers (as opposed to end users), and the primary goals are high performance, interface portability, and expressiveness. GASNet stands for "Global-Address Space Networking" \cite{gasnet-lcpc18}.

In terms of technology, having GASNet as an under-layer means that UPC and UPC++ have to stay synchronized with the GASNet project in order to benefit from the latest version of it. It also means for the user that a GASNet version change requires recompiling of both the UPC or UPC++ compiler and runtime environment, as well as all the programs. The same process is required when new features are implemented in UPC or UPC++: a full recompiling of the whole tool-chain and all the programs.

It is important to specify that UPC++ relies on a newer version of GASNet called GASNet-EX. Also, newest versions of UPC, not used in this paper, rely on GASNet-EX. In \cite{gasnet-lcpc18}, the authors give a detailed presentation of GASNet-EX as well as a short definition of GASNet-EX's "philosophy": "GASNet-EX is the next generation of the GASNet-1 communication system, continuing our commitment
to provide portable, high-performance, production-quality, open-source software. The GASNet-EX upgrade
is being done over the next several years as part of the U.S. Department of Energy’s Exascale Computing
Program (ECP). The GASNet interfaces are being redesigned to accommodate the emerging needs of exascale
supercomputing, providing communication services to a variety of programming models on current and future
HPC architectures. This work builds on fifteen years of lessons learned with GASNet-1, and is informed and
motivated by the evolving needs of distributed runtime systems."
A set of improvements between GASNet and GASNet-EX are also presented in \cite{gasnet-lcpc18}:
\begin{itemize}
    \item Retains GASNet-1’s wide portability (laptops to supercomputers)
    \item Provides backwards compatibility for the dozens of GASNet-1 clients,
    \item including multiple UPC and CAF/Fortran08 compilers
    \item Focus remains on one-sided RMA and Active Messages
    \item Reduces CPU and memory overheads
    \item Improves many-core and multi-threading support
    \item “Immediate mode” injection to avoid stalls due to back-pressure
    \item Explicit handling of local-completion (source buffer lifetime)
    \item New AM interfaces, e.g. to reduce buffer copies between layers
    \item Vector-Index-Strided for non-contiguous point-to-point RMA
    \item Remote Atomics, implemented with NIC offload where available
    \item Subset teams and non-blocking collectives 
\end{itemize}

\section{Results}
\label{results}

In this chapter, we focus on presenting the results obtained with our implementations of UPC++ SpMV (see Section \ref{spmvImplementation}, page \pageref{spmvImplementation}) and UPC++ Heat Equation 2D (see Section \ref{heatEquationImplementation}, page \pageref{heatEquationImplementation}). First we will show that it is possible to get reproducible performance with UPC++. We then discuss the importance of the {\tt BLOCKSIZE} on the data distribution and the obtained performance. Then, we will focus on comparing performance obtained in UPC++ with that of UPC. The UPC results come from our previous study \cite{lagraviere2019performance}.

In this section we use named implementations of UPC++, the correspondence with the implementations presented in section \ref{implementations} is as follows:
\begin{itemize}
    \item "UPC++SpMV Version 1" corresponds to "UPC++ SpMV with Block-Wise Data Transfer between Threads" presented in Section \ref{blockWiseImplementation} (see page \pageref{blockWiseImplementation})
    \item "UPC++SpMV Version 2" corresponds to "UPC++ SpMV with Message condensing and consolidation" presented in Section \ref{cleverComms} (see page \pageref{cleverComms})
    \item "UPC Heat Equation" corresponds to the implementation of UPC the Heat Equation in 2D presented in Section \ref{heatEquationImplementation} (see page \pageref{heatEquationImplementation})
    \item "UPC++ Heat Equation" corresponds to the implementation of the UPC++ Heat Equation in 2D presented in Section \ref{heatEquationImplementation} (see page \pageref{heatEquationImplementation})
\end{itemize}

\subsection{Data distribution: reproducibility and influence on performance}
We focus on the reproducibility of the obtained performance with UPC++ since it is a crucial feature of a new language such as UPC++, both for verifying its usefulness in high performance computing and for ensuring the validity of our results. In addition, being able to expect repeatable performance from UPC++ is a prerequisite for designing a performance model capable of predicting obtainable performance.

\begin{figure}[t]
\centerline{{\includegraphics[width=13.5cm]{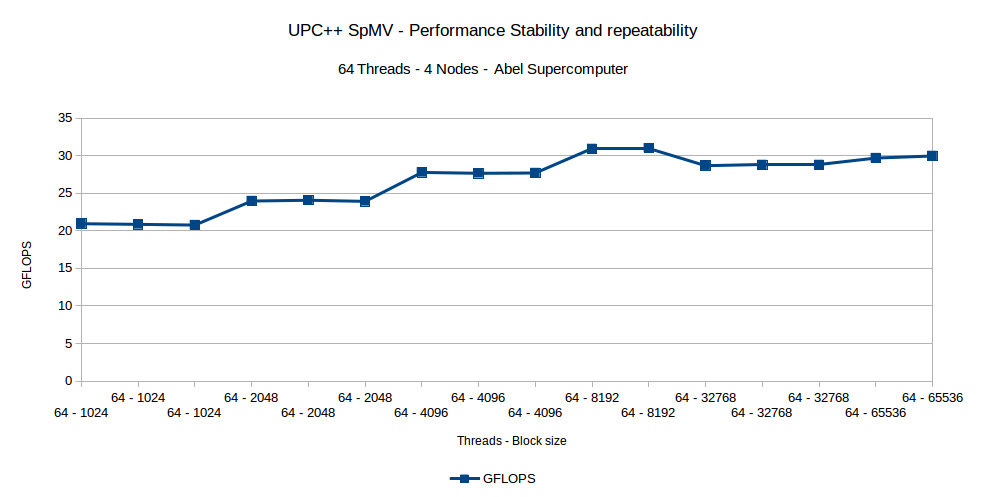}}}
\centerline{{\includegraphics[width=13.5cm]{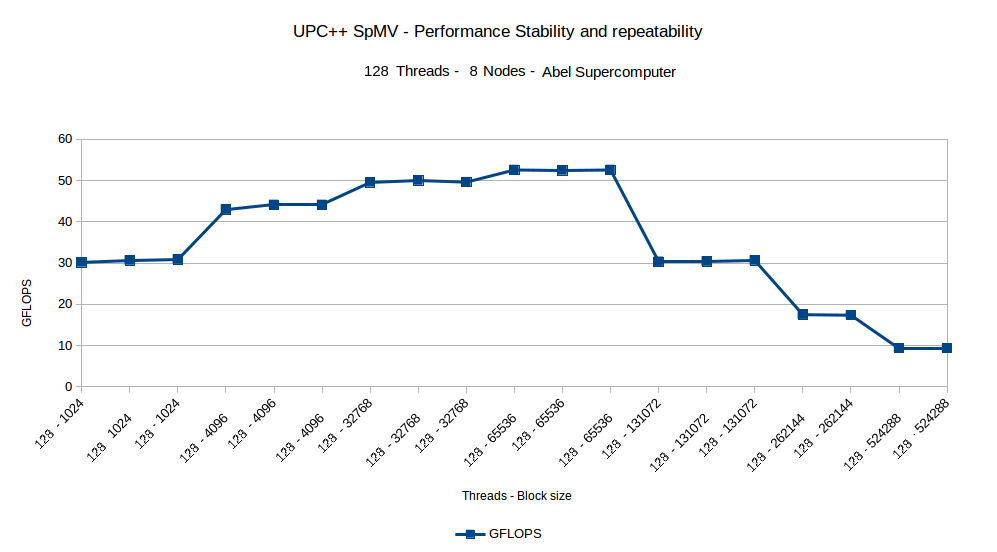}}}
\caption{Performance stability and repeatability and importance of the {\tt BLOCKSIZE} on the obtained performance. Using human heart 3D representation, using 6.8 millions tetrahedrons ($n=6810586$). Using 64 and 128 threads spread
  over 4 and 8 nodes respectively.}
\label{fig:comm_volumes}
\end{figure}

In Figure \ref{fig:comm_volumes}, we show results measured on the Abel supercomputer using UPC++ SpMV version 2 running on two different counts of threads: 64 (4 nodes) and 128 threads (8 nodes), processing an instance representing a healthy human heart composed of 6.8 million tetrahedrons. For each thread count we have used different values of {\tt BLOCKSIZE} in order to investigate two aspects: first whether the {\tt BLOCKSIZE} has a strong impact on performance, and second that when running the same UPC++ SpMV program multiple times with the same parameters {\tt Number of threads - BLOCKSIZE} we obtain stable and repeatable performance.

Figure \ref{fig:comm_volumes} shows both that we obtain performance stability and that the {\tt BLOCKSIZE} has a strong impact on performance. This impact in performance is related to the fact that the chosen {\tt BLOCKSIZE} in our implementations affects directly the data distribution.  Consequently, for this particular data set (6.8 millions tetrahedrons), the observed performance varies significantly depending on the chosen value of {\tt BLOCKSIZE}: for 128 threads (8 nodes) performance go from slightly more than 50 GFLOPS down to slightly less than 10 GFLOPS when using either 65536 or 524288 as {\tt BLOCKSIZE}. For 64 threads, going from a block size of 1024 to 8192 helps to improve the performance, however, when the block size is bigger than 8192 we observe a slight drop in the obtained performance.

\subsection{Scalability and Performance}
\begin{figure}[t]
	\centerline{\includegraphics[width=13.5cm]{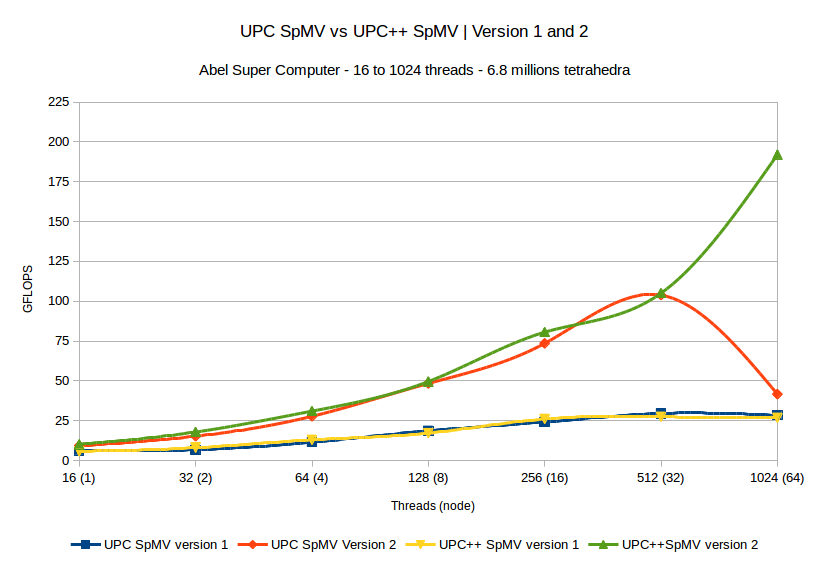}}
	\caption{Multi-node performance for SpMV using Abel Supercomputer (abel.uio.no). Using 1 to 64 nodes, 16 to 1024 threads. Performance are expressed in GFLOPS (higher is better). Using human heart 3D representation, using 6.8 millions tetrahedrons ($n=6810586$). Table \ref{table:spmvAbelResults} contains numerical results represented in this graph.}
	\label{fig:spmvPerformanceAbel}
\end{figure}

\begin{table}[t]
\centering
\begin{tabular}{|c|c|c|c|c|}
\hline
\textbf{Threads (node)} & \textbf{\begin{tabular}[c]{@{}l@{}}UPC SpMV\\ version 1\end{tabular}} & \textbf{\begin{tabular}[c]{@{}l@{}}UPC SpMV\\ version 2\end{tabular}} & \textbf{\begin{tabular}[c]{@{}l@{}}UPC++ SpMV\\ version 1\end{tabular}} & \textbf{\begin{tabular}[c]{@{}l@{}}UPC++SpMV\\ version 2\end{tabular}} \\ \hline
\textbf{16 (1)}         & 6.22                                                                  & 9.20                                                                  & 5.78                                                                    & 10.11                                                                  \\ \hline
\textbf{32 (2)}         & 6.75                                                                  & 15.20                                                                 & 8.05                                                                    & 17.86                                                                  \\ \hline
\textbf{64 (4)}         & 11.46                                                                 & 27.72                                                                 & 12.97                                                                   & 30.99                                                                  \\ \hline
\textbf{128 (8)}        & 18.62                                                                 & 48.35                                                                 & 17.18                                                                   & 50.55                                                                  \\ \hline
\textbf{256 (16)}       & 24.14                                                                 & 73.50                                                                 & 26.10                                                                   & 80.54                                                                  \\ \hline
\textbf{512 (32)}       & 29.57                                                                 & 103.77                                                                & 27.43                                                                   & 105.01                                                                 \\ \hline
\textbf{1024 (64)}      & 28.41                                                                 & 41.61                                                                 & 26.84                                                                   & 192.00                                                                 \\ \hline
\end{tabular}
\caption{Performance results in GFLOPS for UPC++ version 1 and 2 versus UPC SpMV version 1 and 2. Performance obtained on Abel super-computer (described in section \ref{hardware} see page \pageref{hardware}), computing on 16 to 1024 threads (1 to 64 nodes). Using human heart 3D representation, using 6.8 millions tetrahedrons ($n=6810586$). For graphical representation see Figure \ref{fig:spmvPerformanceAbel}.}
\label{table:spmvAbelResults}
\end{table}

\begin{table}[t]
\centering
\begin{tabular}{|c|c|c|c|c|}
\hline
\textbf{Threads (node)} & \textbf{\begin{tabular}[c]{@{}c@{}}UPC SpMV\\ version 1\end{tabular}} & \textbf{\begin{tabular}[c]{@{}c@{}}UPC SpMV\\ version 2\end{tabular}} & \textbf{\begin{tabular}[c]{@{}c@{}}UPC++ SpMV\\ version 1\end{tabular}} & \textbf{\begin{tabular}[c]{@{}c@{}}UPC++SpMV\\ version 2\end{tabular}} \\ \hline
\textbf{16 (1)}         & -                                                                     & -                                                                     & -                                                                       & -                                                                      \\ \hline
\textbf{32 / 16}        & 1.08                                                                  & 1.65                                                                  & 1.39                                                                    & 1.77                                                                   \\ \hline
\textbf{64 / 32}        & 1.70                                                                  & 1.82                                                                  & 1.61                                                                    & 1.73                                                                   \\ \hline
\textbf{128 / 64}       & 1.62                                                                  & 1.74                                                                  & 1.32                                                                    & 1.60                                                                   \\ \hline
\textbf{256 / 128}      & 1.30                                                                  & 1.52                                                                  & 1.52                                                                    & 1.63                                                                   \\ \hline
\textbf{512 / 256}      & 1.22                                                                  & 1.41                                                                  & 1.05                                                                    & 1.30                                                                   \\ \hline
\textbf{1024 / 512}     & 0.96                                                                  & 0.40                                                                  & 0.98                                                                    & 1.83                                                                   \\ \hline
\end{tabular}
\caption{Performance results in speed-up for UPC++ version 1 and 2 versus UPC SpMV version 1 and 2. Performance obtained on Abel supercomputer (described in Section \ref{hardware} see page \pageref{hardware}), computing on 16 to 1024 threads (1 to 64 nodes). Using the human heart 3D representation with 6.8 millions tetrahedrons ($n=6810586$). As an example on how to read this table: For UPC SpMV version 2 the speed-up between using 128 threads and 64 threads is 1.74.}
\label{table:spmvAbelSpeedup}
\end{table}
Concerning multi-node results for UPC SpMV and UPC++ SpMV: Table \ref{table:spmvAbelResults} and Figure \ref{fig:spmvPerformanceAbel} present the obtained performance for UPC and UPC++ SpMV versions 1 and 2 using again the Abel supercomputer. Additionally, we present the obtained speedups for UPC and UPC++ SpMV versions 1 and 2 in Table \ref{table:spmvAbelSpeedup}. We obtain speed-ups strictly superior to 1 for all versions, and all amounts of threads (except for 1024 threads (64 nodes) for UPC SpMV version 1 and 2 and UPC++ version 1). Also, the obtained speed-up on the Abel supercomputer for all versions of UPC and UPC++ SpMV is always strictly inferior to 2.00.

Based on these results we can say that we achieve strong scaling in UPC SpMV and UPC++ SpMV version 1 and 2 using the Abel supercomputer for the multi-node experiments (except for UPC version 1 and 2 and UPC++ version 1 for 1024 threads (64 nodes)). The obtained speedups can be called  \emph{strong scaling} in the sense that for the same total amount of data (fixed problem size) we get higher performance when using more processing elements from 16 to 512 cores (1 to 32 nodes), when using 1024 threads only UPC++ version 2 continues to scale up: 1.83 speed-up (see Table \ref{table:spmvAbelSpeedup}, page \pageref{table:spmvAbelSpeedup}) between UPC++ version 2 using 1024 threads (64 nodes) compared to performance obtained with UPC++ version 2 using 512 threads (32 nodes).

In Figure \ref{fig:spmvPerformanceAbel}, it is clear that message condensing and consolidation (which corresponds to UPC SpMV version 2 and UPC++ SpMV version 2) delivers far better performance than block wise data transfer (which corresponds to UPC version 1 and UPC++ version 1).

This is to be expected as the versions 2 of UPC and UPC++ SpMV use a far more sophisticated communication pattern, leading to messages sent between threads/cores/nodes of a size that corresponds exactly to the amount of data we need to transfer. Whereas in the block-wise data transfer versions (UPC SpMV and UPC++ SpMV versions 1), messages are always of size {\tt BLOCKSIZE} even if only 1 element (C-type "double", 8 bytes on 64 bits architecture) needs to be sent.
Thus, communication containing only one element that is needed will be of size $ 1 \times 8(bytes) $ in UPC and UPC++ SpMV versions 2, and of size $ 1 \times BLOCKSIZE \times 8 (bytes) $ in UPC and UPC++ SpMV versions 1. As the communication pattern is the only main difference in UPC and UPC++ versions 1 and 2, it is possible to say that the communication pattern is the main source of difference in performance. In addition, we have shown in our previous study that this performance can be predicted using our performance model \cite{lagraviere2019performance}, and we show in this study that on multi-node the heaviest factor influencing performance is the inter-node communication through the Infiniband network.

\begin{figure}[t]
	\centerline{\includegraphics[width=13.5cm]{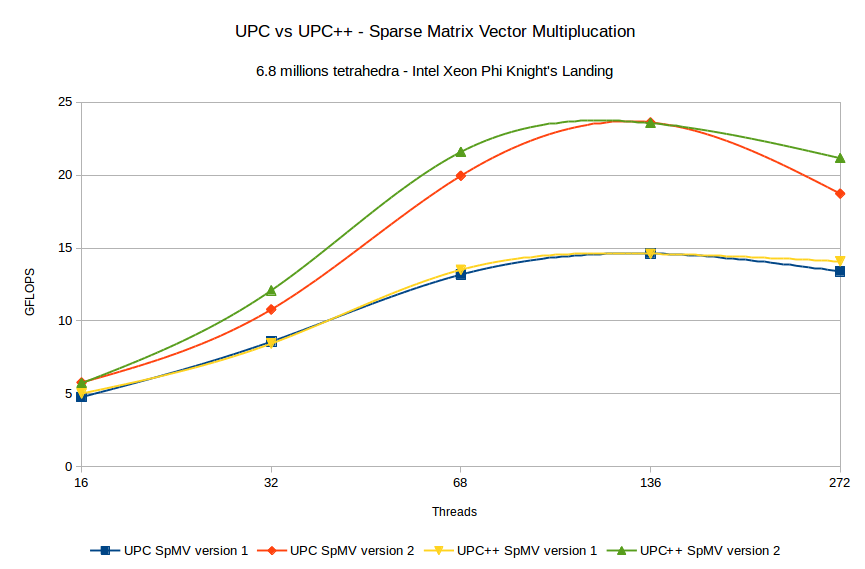}}
	\caption{Single-node performance for SpMV using Intel Xeon Phi Knight's Landing (described in section \ref{hardware}, see page \pageref{hardware}). Using human heart 3D representation, using 6.8 millions tetrahedrons ($n=6810586$).}
	\label{fig:spmvPerformanceKnl}
\end{figure}

\begin{table}[b]
\centering
\begin{tabular}{|c|c|c|c|c|}
\hline
\textbf{Threads} & \textbf{\begin{tabular}[c]{@{}c@{}}UPC SpMV\\ version 1\end{tabular}} & \textbf{\begin{tabular}[c]{@{}c@{}}UPC SpMV\\ version 2\end{tabular}} & \textbf{\begin{tabular}[c]{@{}c@{}}UPC++ SpMV\\ version 1\end{tabular}} & \textbf{\begin{tabular}[c]{@{}c@{}}UPC++ SpMV\\ version 2\end{tabular}} \\ \hline
\textbf{16}      & 4.76                                                                  & 5.76                                                                  & 5.00                                                                    & 5.71                                                                    \\ \hline
\textbf{32}      & 8.57                                                                  & 10.77                                                                 & 8.43                                                                    & 12.08                                                                   \\ \hline
\textbf{68}      & 13.17                                                                 & 19.95                                                                 & 13.50                                                                   & 21.59                                                                   \\ \hline
\textbf{136}     & 14.62                                                                 & 23.61                                                                 & 14.60                                                                   & 23.57                                                                   \\ \hline
\textbf{272}     & 13.39                                                                 & 18.72                                                                 & 14.08                                                                   & 21.16                                                                   \\ \hline
\end{tabular}

\caption{Performance results in GFLOPS for UPC++ version 1 and 2 versus UPC SpMV version 1 and 2. Performance obtained on Intel Xeon Phi Knight's Landing many-core processor (Described in section \ref{hardware} see page \pageref{hardware}), computing on 16 to 272 threads. Using human heart 3D representation, using 6.8 millions tetrahedrons ($n=6810586$). For graphical representation see Figure \ref{fig:spmvPerformanceKnl}. }
\label{table:spmvKnlResults}
\end{table}

Table \ref{table:spmvKnlResults} (Page \pageref{table:spmvKnlResults}) and Figure \ref{fig:spmvPerformanceKnl} show the obtained performance for UPC SpMV and UPC++ SpMV versions 1 and 2 on the Intel Xeon Phi Knight's Landing (as presented in Section \ref{hardware}, page \pageref{hardware}). 
Globally, UPC and UPC++ SpMV in versions 2 perform better than UPC and UPC++ SpMV versions 1. As explained for the multi-node results, the main difference between these two versions is the communication pattern. In this single-node context running on a many-core architecture, the difference in the communication pattern has a huge influence on performance. As shown earlier in this section and in section \ref{spmvImplementation} (page \pageref{spmvImplementation}) and in \cite{lagraviere2019performance}, block-wise data transfer, as implemented in UPC and UPC++ SpMV versions 1, generate far more data traffic than message condensing and consolidation, as implemented in UPC and UPC++ versions 2. This difference in the amount of data exchanged between cores on a many-core such as Intel Xeon Phi Knight's Landing architecture is large enough to lead to bottlenecks between the processing elements and the MCDRAM (see Section \ref{hardware}, page \pageref{hardware}).

\begin{table}[t]
\centering
\begin{tabular}{|c|c|c|}
\hline
\textbf{Threads (node)} & \textbf{UPC Heat Equation} & \textbf{UPC++ Heat Equation} \\ \hline
\textbf{16 (1)}         & 123.04                     & 123.03                       \\ \hline
\textbf{32 (2)}         & 64.99                      & 64.00                           \\ \hline
\textbf{64 (4)}         & 31.05                      & 33.00                           \\ \hline
\textbf{128 (8)}        & 15.60                       & 14.20                         \\ \hline
\textbf{256 (16)}       & 7.88                       & 8.90                          \\ \hline
\textbf{512 (32)}       & 3.99                       & 3.85                         \\ \hline
\textbf{1024 (64)}      & 3.26                   & 0.52                         \\ \hline
\end{tabular}
\caption{Performance results in seconds for UPC++ Heat Equation 2D versus UPC Heat Equation 2D. Domain size {\tt 20000x20000}, 1000 iterations. Performance obtained on Abel super-computer (described in section \ref{hardware} see page \pageref{hardware}), computing on 16 to 1024 threads (1 to 64 nodes). For graphical representation see Figure \ref{fig:heatEquationPerformanceAbel}.}
\label{table:heatEquationAbelResults}
\end{table}
\begin{figure}[t]
	\centerline{\includegraphics[width=12cm]{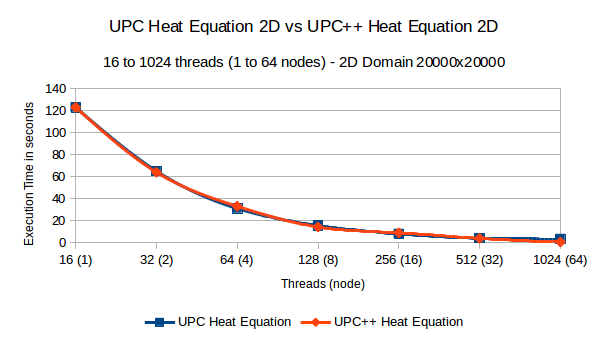}}
	\caption{Multi-node performance for Heat Equation 2D, domain size 20000x20000, 1000 iterations using Abel Supercomputer (abel.uio.no). Using 1 to 64 nodes, 16 to 1024 threads. Performance are expressed in seconds (execution time) (lower is better). Table \ref{table:heatEquationAbelResults} contains numerical results represented in this graph.}
	\label{fig:heatEquationPerformanceAbel}
\end{figure}

Additionally, since the Intel Knight's Landing's hardware architecture uses four-way \emph{HyperThreading}, i.e.~each of the 68 physical cores appears to the operating system as four logical cores. Since the logical cores do not add computational resources,
we cannot expect linear scaling from using them, as observed in \cite{tau2002empirical}. This explains why when using 136 cores we only get a slight increase in performance. When using 272 cores we observe a performance drop for all versions of UPC and UPC++ SpMV.

Table \ref{table:heatEquationAbelResults} and Figure \ref{fig:heatEquationPerformanceAbel}, present the obtained results for our implementations of UPC and UPC++ Heat Equation using a 20000x20000 domain running on Abel supercomputer using 16 to 1024 cores (1 to 64 nodes). Both the UPC and UPC++ Heat Equation versions adopt a similar way of communicating data and a similar computation kernel, as presented in Section \ref{heatEquationImplementation}, page \ref{heatEquationImplementation}. With that in mind we observe that UPC++ performs as well as UPC. In this case, the data per processor is fixed independently of the amount of cores that is used, thus, as for our implementations of SpMV, we can observe that we achieve weak scaling in both UPC and UPC++, at least from 1 to 32 nodes.

The obtained performance is extremely close between the UPC and UPC++ implementations of the Heat Equation, from 1 to 32 nodes. However when using 64 nodes (1024 cores) UPC++ has a strong advantage over UPC. We observed this behavior of UPC having trouble to perform with more than 512 threads before \cite{lagraviere2019performance,lagraviere2016performance}.

\section{Discussion}
\label{discussion}
In Section \ref{implementations} (page \pageref{implementations}), we have presented our  UPC and UPC++ implementations of SpMV and the Heat Equation in 2D. It is important to note the significant differences in both programming model and consequences in terms of code between UPC and UPC++. For instance, the way the shared arrays are accessed in UPC is done relatively transparently in the sense that a shared array of size {\tt N} will be accessible with indices ranging from {\tt 0} to {\tt N-1}. However, in UPC++ the same array will be accessible in a 2-dimensional way, the first dimension explicitly points at the thread ID, the second dimension corresponds to a position in the array for the portion allocated to the chosen thread. 

In other words, UPC++ does not try to maintain the same ease of programming at the expense of performance as UPC. As shown in previous work \cite{lagraviere2019performance}, accessing shared arrays in UPC can be extremely costly as it generates \emph{hidden}  communication: hidden in two senses, first the programmer does not explicitly implement the communication and the developer has no way to know whether this communication will be between cores, sockets or nodes. In other words, in UPC communication can be implicit and yield extremely bad performance whereas in UPC++ all communication is done explicitly. However, UPC++ performs also communication without having the developer know whether the communication is between cores, sockets or nodes (unless \emph{teams} are used). This simply means that the UPC and/or UPC++ programmer has to pay attention to both data distribution and the use of communication whether it is implicit or explicit.

This leads to an increase in the \emph{effort in programming} that UPC++ developers have to deploy in order to get satisfying performance on single-node, multi-node or many-core hardware architectures with UPC++ it is no longer possible to claim that this implementation of the PGAS programming model means a better \emph{ease of programming} compared to the de-facto standard MPI+OpenMP. However, by offering a lot of features related to one-sided asynchronous communication, the use of C++ and \emph{Promises} and \emph{Futures} in order to enforce the use of asynchronous programming, UPC++ represents an alternative to MPI+OpeMP. Moreover, UPC++ offers these functionalities and yields performance as we explain below and as we have shown in this paper.

Our goal was also to verify and show whether UPC++ can run and yield good performance on
multi-node and many-core architectures and whether this performance is comparable to previously obtained performance in UPC. Thus, in Section \ref{results} (page \pageref{results}), we have presented the obtained performance for our implementations of UPC and UPC++ SpMV on both multi-node and many-core hardware architecture and we have presented the results for our implementations of UPC and UPC++ Heat Equation on multi-node. Globally UPC++ perform as well as UPC for both considered kernels. This is encouraging as a basis for future work using more advanced UPC++'s features such as asynchronous remote procedure calls and implementation of communication overlapping with computation.

\clearpage
\section{Conclusion}
\label{conclusion}
In this study, we have investigated and provided insights into the UPC++ programming model and its performance. We have measured the computational performance of two kernels (SpMV and Heat Equation 2D) both on a multi-node supercomputer and a many-core processor. 
In addition, we have compared UPC++ to UPC by running identical kernels. We have shown that both on multi-node and many-core architecture UPC++ can compete with UPC and yields  better performance on the largest amounts of cores and nodes that we have used (1024 cores, 64 nodes).

We have provided a detailed view of the difference of programming styles between UPC and UPC++ and the consequences in terms of code. UPC and UPC++, despite their similar names, are two extremely different implementations of the PGAS programming model, and in that sense, switching from UPC to UPC++ should not be considered lightly.

For future studies, there are many tracks to investigate, such as using more UPC++ features such as asynchronous remote procedure calls, or using UPC++ in conjunction with CUDA, and using UPC++ to implement different kind of kernels as we have focused our studies mainly on memory-bound and communication-bound problems.

\clearpage
\bibliographystyle{abbrv}
\bibliography{paper}

\begin{thebibliography}{10}

\bibitem{abel_cluster}
{The Abel computer cluster}.
\newblock \url{www.uio.no/english/services/it/research/hpc/abel/}, 2018.

\bibitem{Almasi2011}
G.~Almasi.
\newblock {PGAS} (partitioned global address space) languages.
\newblock In D.~Padua, editor, {\em Encyclopedia of Parallel Computing}, pages
  1539--1545. Springer, 2011.

\bibitem{bachan2019upc++}
J.~Bachan, S.~B. Baden, S.~Hofmeyr, M.~Jacquelin, A.~Kamil, D.~Bonachea, P.~H.
  Hargrove, and H.~Ahmed.
\newblock {UPC++": A High-Performance Communication Framework for Asynchronous
  Computation}.
\newblock In {\em 2019 IEEE International Parallel and Distributed Processing
  Symposium (IPDPS)}, pages 963--973. IEEE, 2019.

\bibitem{Berkeley_UPC}
{Berkeley UPC -- Unified Parallel C}.
\newblock \url{upc.lbl.gov}, 2018.

\bibitem{gasnet-lcpc18}
D.~Bonachea and P.~H. Hargrove.
\newblock {GASNet-EX: A High-Performance, Portable Communication Library for
  Exascale}.
\newblock In {\em Proceedings of Languages and Compilers for Parallel Computing
  (LCPC'18)}, volume 11882 of {\em Lecture Notes in Computer Science}. Springer
  International Publishing, October 2018.

\bibitem{UPCPointOfOrigin}
W.~W. Carlson, J.~M. Draper, D.~E. Culler, K.~Yelick, E.~Brooks, and K.~Warren.
\newblock {Introduction to UPC and language specification}.
\newblock Technical report, Technical Report CCS-TR-99-157, IDA Center for
  Computing Sciences, 1999.

\bibitem{intelKnlMemoryStuff}
{Colfax Research}.
\newblock {MCDRAM as High-Bandwidth Memory (HBM) in Knights Landing Processors:
  Developer’s Guide}.
\newblock \url{https://colfaxresearch.com/knl-mcdram/}, 2017.
\newblock [Online; accessed 20-November-2019].

\bibitem{1263470}
D.~E. Culler, A.~Dusseau, S.~C. Goldstein, A.~Krishnamurthy, S.~Lumetta,
  T.~{von Eicken}, and K.~Yelick.
\newblock Parallel programming in {Split-C}.
\newblock In {\em Supercomputing '93. Proceedings}, pages 262--273, Nov 1993.

\bibitem{davies1970domain}
A.~Davies and J.~Mushtaq.
\newblock The domain decomposition boundary element method on a network of
  transputers.
\newblock {\em WIT Transactions on Modelling and Simulation}, 15, 1970.

\bibitem{PGAS_survey}
M.~{de Wael}, S.~Marr, B.~{de Fraine}, T.~{van Cutsem}, and W.~{de Meuter}.
\newblock Partitioned global address space languages.
\newblock {\em ACM Computing Surveys}, 47(4), 2015.

\bibitem{chapelPointOfOrigin}
S.~J. Deitz, B.~L. Chamberlain, and M.~B. Hribar.
\newblock {Chapel: Cascade High-Productivity Language An Overview of the Chapel
  Parallel Programming Model}.
\newblock {\em Cray User Group}, 2006.

\bibitem{Grimes1979}
R.~Grimes, D.~Kincaid, and D.~Young.
\newblock {ITPACK 2.0 User’s Guide}.
\newblock Technical Report CNA-150, Center for Numerical Analysis, University
  of Texas, 1979.

\bibitem{lagraviere2019performance}
J.~Lagravi{\`e}re, J.~Langguth, M.~Prugger, L.~Einkemmer, P.~H. Ha, and X.~Cai.
\newblock {Performance Optimization and Modeling of Fine-Grained Irregular
  Communication in {UPC}}.
\newblock {\em Scientific Programming}, Article ID 6825728, 2019.

\bibitem{lagraviere2016performance}
J.~Lagraviere, J.~Langguth, M.~Sourouri, P.~H. Ha, and X.~Cai.
\newblock {On the performance and energy efficiency of the pgas programming
  model on multicore architectures}.
\newblock In {\em 2016 International Conference on High Performance Computing
  \& Simulation (HPCS)}, pages 800--807. IEEE, 2016.

\bibitem{langguth2017porting}
J.~Langguth, C.~Jarvis, and X.~Cai.
\newblock Porting tissue-scale cardiac simulations to the knights landing
  platform.
\newblock In {\em International Conference on High Performance Computing},
  pages 376--388. Springer, 2017.

\bibitem{PGASAsAnAlternative}
D.~A. Mall{\'o}n, G.~L. Taboada, C.~Teijeiro, J.~Touri{\~{n}}o, B.~B. Fraguela,
  A.~G{\'o}mez, R.~Doallo, and J.~C. Mouri{\~{n}}o.
\newblock {Performance Evaluation of MPI, UPC and OpenMP on Multicore
  Architectures}.
\newblock In M.~Ropo, J.~Westerholm, and J.~Dongarra, editors, {\em Recent
  Advances in Parallel Virtual Machine and Message Passing Interface}, pages
  174--184, Berlin, Heidelberg, 2009. Springer Berlin Heidelberg.

\bibitem{martin2018mate}
S.~M. Martin.
\newblock {\em MATE, a Unified Model for Communication-Tolerant Scientific
  Applications}.
\newblock PhD thesis, UC San Diego, 2018.

\bibitem{coArrayFortranPointOfOrigin}
R.~W. Numrich and J.~Reid.
\newblock {Co-Array Fortran for parallel programming}.
\newblock In {\em ACM Sigplan Fortran Forum}, volume~17, pages 1--31. ACM,
  1998.

\bibitem{PGAS_site}
{PGAS - Partitioned Global Address Space}.
\newblock \url{http://www.pgas.org}, 2016.

\bibitem{PGAS_course2015}
R.~Rabenseifner.
\newblock {Introduction to Unified Parallel C (UPC) and Co-array Fortran
  (CAF)}, 2015.
\newblock Short course at HLRS, University of Stuttgart, April 23-April 24.

\bibitem{tau2002empirical}
R.~A. Tau~Leng, J.~Hsieh, V.~Mashayekhi, and R.~Rooholamini.
\newblock An empirical study of hyper-threading in high performance computing
  clusters.
\newblock {\em Linux HPC Revolution}, 45, 2002.

\bibitem{upc++OfficialWebpage}
UPC++.
\newblock {UPC++ Official Website}.
\newblock \url{bitbucket.org/berkeleylab/upcxx/}.
\newblock Online; accessed 28-October-2019.

\bibitem{upc++programmersGuide}
UPC++.
\newblock {UPC++ Programmer's Guide, Revision 2019.9.0}.
\newblock
  \url{https://bitbucket.org/berkeleylab/upcxx/downloads/upcxx-guide-2019.9.0.pdf}.
\newblock Online; accessed 28-October-2019.

\bibitem{upcxxchangelog}
UPC++.
\newblock {UPC++ Version Changelog}.
\newblock \url{bitbucket.org/berkeleylab/upcxx/wiki/ChangeLog}.
\newblock Online; accessed 28-October-2019.

\bibitem{Futures_and_promises}
Wikipedia.
\newblock {{Futures and promises} --- {W}ikipedia{,} The Free Encyclopedia}.
\newblock \url{en.wikipedia.org/wiki/Futures_and_promises}, 2019.
\newblock [Online; accessed 28-October-2019].

\bibitem{X10}
{{X10}}.
\newblock {{Performance and Productivity at Scale}}.
\newblock \url{x10-lang.org/}, 2019.
\newblock [Online; accessed 11-December-2019].

\bibitem{6877339}
Y.~{Zheng}, A.~{Kamil}, M.~B. {Driscoll}, H.~{Shan}, and K.~{Yelick}.
\newblock {UPC++: A PGAS Extension for C++}.
\newblock In {\em 2014 IEEE 28th International Parallel and Distributed
  Processing Symposium}, pages 1105--1114, May 2014.

\end{thebibliography}


\begin{thebibliography}{1}

\bibitem{IEEEhowto:kopka}
H.~Kopka and P.~W. Daly, \emph{A Guide to \LaTeX}, 3rd~ed.\hskip 1em plus
  0.5em minus 0.4em\relax Harlow, England: Addison-Wesley, 1999.

\end{thebibliography}

{\bf Conflicts of interest}:
The authors declare that there is no conflict of interest regarding
the publication of this paper.

{\bf Acknowledgements}:
This work was performed on hardware resources provided by UNINETT
Sigma2 -- the National Infrastructure for High Performance Computing
and Data Storage in Norway, via Project NN2849K. 
The work was supported by the European Union's Horizon 2020 research and innovation programme under grant agreement No.~671657, the European Union Seventh Framework Programme (grant No.~611183) and the Research Council of
Norway (grants No.~231746/F20, No.~214113/F20 \& No.~251186/F20).

\end{document}